# Science Merit Function for the *KEPLER* Mission

**William J. Borucki***

NASA Ames Research Center, Mail Stop 244-30, Moffett Field, CA, USA, 94035-0001

**Abstract**

The *Kepler* Mission was a NASA Discovery-class mission designed to continuously monitor the brightness of at least 100,000 stars to determine the frequency of Earth-size and larger planets orbiting other stars. Once the Kepler proposal was chosen for a flight opportunity, it was necessary to optimize the design to accomplish the ambitious goals specified in the proposal and still stay within the available resources. To maximize the science return from the mission, a merit function (MF) was constructed that relates the science value (as determined by the PI and the Science Team) to the chosen mission characteristics and to models of the planetary and stellar systems. This MF served several purposes; prediction of the science results of the proposed mission, effects of varying the values of the mission parameters to increase the science product or to reduce the mission costs, and assessment of risks. The Merit Function was also valuable for the purposes of advocating the Mission by illustrating its expected capability. Later, it was used to keep management informed of the changing mission capability as trade-offs and mission down-sizing occurred.

The MF consisted of models of the stellar environment, assumed exoplanet characteristics and distributions, parameter values for the mission point-design, and equations that related the science value to the predicted number and distributions of detected exoplanets. A description of the MF model and representative results are presented.

- William J. Borucki, William.J.Borucki@nasa.gov

# 1. Introduction

The *Kepler* Mission began as a PI-led NASA Discovery-class mission (#10) that was designed to explore the structure and diversity of planetary systems. Specifically, the Mission was to survey the extended solar neighborhood to detect and characterize hundreds of terrestrial and larger planets in or near the habitable zone (HZ). *Kepler* completed its prime mission in 2013 as a NASA strategic mission. Once the *Kepler* Concept Study Report (CSR)[1] was chosen for a flight opportunity, it was necessary to optimize the design to maximize the scientific productivity given the available resources.

At the time of the first proposal (1992) to the Discovery Announcement of Opportunity (AO), no planets orbiting stars like the Sun had been discovered and no previous mission had been designed to use the photometric method to discover Earth-size planets. After proposing for the 1992, 1994, 1996, and 2000 Discovery AOs, the *Kepler* Mission was accepted for flight development in 2001. Comprehensive engineering studies were started to fully characterize the mission parameters while minimizing the resources necessary to accomplish the mission requirements. Because no previous mission had been designed to do a high-precision (i.e., 10 parts-per-million (ppm) for bright, quiet stars) photometric search of thousands of stars to find planets, there was little history or experience to guide the development of such a mission. Fortunately, ground-based observatories that conducted searches for exoplanets by the photometric method had been constructed during the period between 1994 and 2000 and some experience gained from operating the observatories[1-5].

At the beginning of Mission, the number and distribution of terrestrial planets expected in extra-solar planetary systems (exoplanets) were unknown. Theories of planetary system formation were not well enough developed to make useful predictions of their sizes, occurrence frequencies, or distributions. It was commonly assumed that other planetary systems would be similar to the Solar System with small rocky planets close to the star and giant planets at much larger distances. Early models of planetary system formation[6-9] predicted the formation of many planets smaller than that of the Earth and few larger than the Earth for values of the semi-major axes near 1 AU. Contrary to such expectations, the early ground-based observations[10,f,11] showed the presence of giant planets in short period orbits. These observations were explained as planets that formed beyond the ice-line and later migrated inward toward their parent stars. During their inward spiral it was expected that they would have swept away most of the small rocky planets. Because the mechanism to stop inward spiral before the planet fell into its star was unknown, it was possible that many stars had been swept clean of planets. Consequently the *Kepler* Mission was designed to characterize planetary systems for several possible situations; i.e., 1) like our own, 2) with only giant planets in inner orbits, 3) without any planets, and 4) with many small planets near 1 Astronomical Unit (AU).

---

[11] The CSR was the required study following the down-select of the Year-2000 Kepler proposal (NASA, Announcement of Opportunity Discovery Program 2000, A.O. 00-OSS-02, 2000.6). Authorization to proceed with mission development was based on a review of the CSR.

Before the *Kepler* Mission proposal could be considered for implementation, quantitative predictions for the expected results and a discussion of the risks were required. Therefore estimates of planetary sizes and distributions were necessary prior to mission development to provide estimates of the mission science return. Because theories predicted that the formation of stars implied the concurrent formation of planetary systems, the Merit Function model assumed that most stars would have planets and that a sufficiently large sample of stars would be observed such that a null result would be significant. In particular it was assumed that if detections of 50 or more Earth-size planets in the HZ of solar-like stars could be expected, then the Mission goals would be satisfied. Based on a geometric probability that the orbit of a planet near the HZ was aligned with our line of sight, and a working estimate that 10% of stars would have such planets, the Mission would need to monitor at least 100,000 stars. Thus the original proposal assumed 100,000 stars would be monitored with a sensitivity to detect Earth-size planets orbiting solar-like stars and that the mission duration would be four years. In the successor to the proposal (i.e., the Concept Study Report), the number of target stars was increased to 170,000.

Upon selection for mission development, it was imperative to develop a Merit Function (MF) to evaluate the proposed mission capability to produce the required information under a variety of situations. The MF needed to connect the accomplishment of mission science goals to the mission design parameters and to assumed values of astrophysical quantities and risks. This MF served several purposes; it related the specific science goals with the specific instrument and mission performances, related the mission performance with respect to the "mission base and floor" during mission development, it evaluated trade-offs, and it considered risks. When trade-offs were made, various options were explored and those that did most damage to the science value were identified and rejected. In a proactive approach, the derivative of the science results to changes in mission parameters were estimated, and changes were made that enhanced the science when that could be done for minimal increases in resources. An especially valuable use of the MF was to predict the expected results as a function of assumptions about extrasolar planet frequencies, planet size and distributions, and associations to star characteristics. Risks were assessed by running "What if" scenarios to determine the expected effects of unknowns and to various levels of noise sources from the instrument and from stellar variability. The ability to estimate the effects of unknowns and mission-design changes on the expected science product was particularly important for an exploratory mission where little was known about the situation to be investigated. Ref. 12 provides an example of a pre-launch prediction for the *Kepler* Mission.

The MF was an algorithm designed to produce numerical values for the science return given a set of inputs representing an instrument/spacecraft point-design, estimates of the stellar structure of galaxy, exoplanet size and frequency distributions, and stellar and instrument noise.  For example, increasing the mission duration affects the number of expected detections and the types of stars that can be searched for planets in their HZs. The MF included tables of stellar distributions of brightness, size, temperature, and mass. A model of the expected performance for the instrument point design was also included.

Calculations based on these inputs estimated the Signal-to-Noise-Ratio (SNR) for each star and thus the number of planets that could be detected in various orbits. A rating system was included that assigned values to the minimum size of a planet that could be detected as a function of the semi-major axis. In turn, the selection of the values for the semi-axes determined the number of transits observed for a given stellar mass. The model considered two separate aspects of the Mission goals; assessing the capability of the Mission to detect small planets (i.e., planet radius (Rp) < 2 Earth-radius ($R_\oplus$)) in the HZ, and the capability to determine the structure of planetary systems by finding small planets at a range of semi-major axes that were near, but not in the HZ. Because the structures of other planetary systems were unknown, the model assumed that other planetary systems were similar to the Solar System in having 3 planets in or near the HZ.

Two complementary algorithms (labeled "Model 1" and "Model 2") were developed and their scores were combined to provide a single value for the MF. The first algorithm computed a score based entirely on the number and size of detectable planets in the HZ of each star. The second algorithm computed a score for Earth-size planets placed in several orbits near, but not in, the HZ. The 9 values of the semi-axis ranged from 0.05 AU to the orbital distance of Mars (1.5 AU). This range covered positions in and near the HZ for star types from M- to G-dwarfs. Together, the two algorithms addressed the goal of determining the occurrence frequency of small planets in the HZ and the goal of exploring the diversity of planetary systems.

The results of the MF were found to be valuable for the purposes of illustrating the Mission capability before it was selected for implementation, for understanding risk, and for keeping mission managers informed of the changing status of the Mission capability during its development.

In this paper, the science goals and mission overview are covered in §1. A short overview of the Mission approach is presented in §2 while the structure of MF is discussed in §3. The expected results from mission are given in §4 and some factors controlling the results are presented in §5. Results from risk studies are considered in §6. A brief comparison of model predictions and Mission results is presented in §7. A summary with a discussion of strengths and weaknesses of the MF approach is presented §8.

## 2. Science goals and overview of mission approach

The scientific goal of the *Kepler* Mission (as stated in the Concept Study Report that was the successor to the Year-2000 mission proposal[13]) was to explore the structure and diversity of planetary systems with special emphasis on determining the frequency of Earth-size planets in the HZ of solar-like stars. This was achieved by surveying approximately 170,000 stars to:

1. Determine the frequency of 0.8 $R_{\oplus}$ and larger planets in or near the HZ of a wide variety of spectral types of stars;
2. Determine the distributions of sizes and orbital semi-major axes of these planets;
3. Estimate the frequency of planets orbiting multiple-star systems;
4. Determine the distributions of semi-major axis, eccentricity, albedo, size, mass, and density of short period giant planets;
5. Identify additional members of each photometrically-discovered planetary system using complementary techniques; and
6. Determine the properties of those stars that harbor planetary systems.

A photometer was designed to be placed in heliocentric orbit to observe the periodic dimming in starlight caused by planetary transits. The photometer consisted of a 0.95 meter, Schmidt-type telescope with a 113 square-degree (sq-deg) field-of-view (FOV). During the 4 years of its operation it monitored the individual brightness of over 170,000 stars imaged onto a large array of CCD detectors[13]. Several thousands of the targets were changed on a quarterly basis to accommodate asteroseismology and guest-investigator studies. Thus, over 190,000 stars were observed during the mission duration prior to the failure of two of its reaction wheels.

The time-series brightness observations measured the depth and time of the transits. These values were used to estimate;

- planet size from the brightness change and from independent measurement of the stellar radius,
- orbital period from the time between transits,
- semi-major axes based on the orbital period, the mass of the star, and *Kepler*'s 3$^{rd}$ Law;
- whether the planet was in the HZ based on the heat flux incident on the planet calculated from the distance to the host star and on the stellar size and temperature.

The HZ was defined as the region around the host star where water on the surface of a rocky planet could be in liquid form and was considered the region most likely to be conducive to the evolution of life.

To help distinguish planetary transits from statistical fluctuations in the data and from astrophysical phenomena that mimic transits, at least three transits were required for a valid detection. Consequently the Mission design required a mission duration of at least three times the longest orbital period sought. Analysis using the MF indicated that the Mission duration should be four years to enhance the detectability of Earth-size planets transiting stars as large as the Sun.

Figure 1 illustrates the general layout of the instrument. The detector array is at prime focus and is cooled by heat pipes that carry the heat to a radiator in the shadow of the spacecraft. The low-level electronics are placed immediately behind the focal array. Because of the very large FOV, the focal surface is strongly curved. To focus the images on the flat charge-coupled detectors (CCDs), each CCD is covered with a field-flattening sapphire lens. A four-vane spider supports the focal plane and its electronics and contains the power- and signal-cables and the heat pipes.

The spacecraft bus encloses the base of the photometer and provides solar panels for power and contains the communication, navigation, and power equipment (Figure 2). Several antennas with different frequency coverage and gain patterns are present for uplink commanding and for data downlink. Approximately 9 Gigabytes (GB) of science data were transferred to the ground every month when contact was made with the NASA Deep-Space Network. More extensive information and references can be found in Refs. 13 & 14.

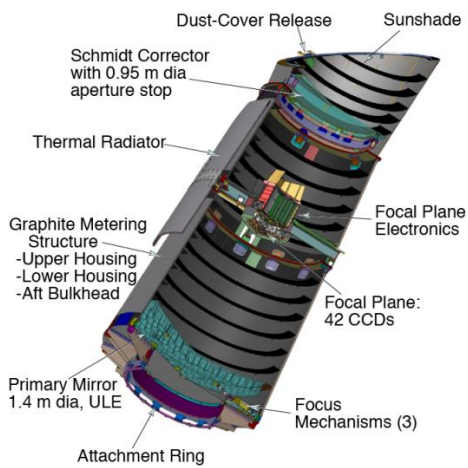

**Figure 1. Schematic diagram of the point design of the *Kepler* instrument.**

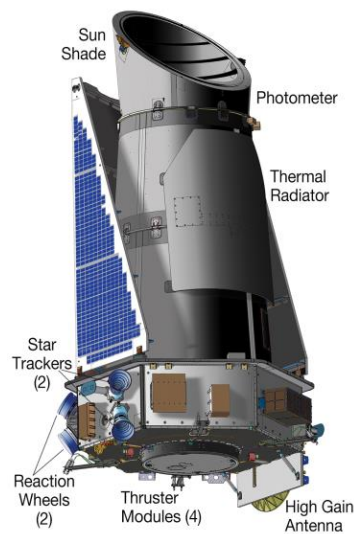

**Figure 2. Integrated spacecraft and photometer.** Cartoons of the instrument and the spacecraft from Ball Aerospace and Technology Corporation (BATC) in Boulder, Colorado. BATC built the instrument and spacecraft. JPL managed the Mission after the selection of the mission. NASA Ames managed it during mission operations.

The point design for the Mission assumed;
  i)   113 sq-deg FOV,
  ii)  0.95 m aperture with a 30% blockage due to the detectors situated at prime focus
  iii) 170,000 stars monitored simultaneously
  iv)  94% duty cycle for science observations after commissioning
  v)   Mission duration of 4 years

## 3. Merit Function description

The MF had 6 parts;
  i)   Model of the stellar characteristics and distribution of apparent brightness
  ii)  Models of the exoplanet characteristics and distribution
  iii) Spacecraft and instrument point design
  iv)  Assignment of values for Science Products
  v)   Algorithms to calculate the expected science products
  vi)  Presentation and interpretation of the results

Model values for stellar masses, sizes, and temperatures

Tables of input values for stellar masses and sizes were obtained from Ref. 15. The distributions for areal density of stars as functions of spectral type and apparent magnitude were based on Besancon Model[16]. To maximize the number of stars observed, the FOV was chosen to be near the galactic plane. The stellar model included the stellar mass, size, and temperature for spectral types (SpT) from A5 to M7. However a case was considered where stars earlier than F5 were omitted because they would be too large to produce useful transit amplitudes or were too rare or had spectra that were inappropriate for planet confirmation by radial-velocity (RV) observations. Similarly, stars later than M7 were considered to be too dim to contribute meaningfully to the results.

Calculations showed that the power spectrum of stellar variability, even at a level of several ppm, influenced the minimum size and orbital characteristics of the planets that could be detected.  The noise level contribution was a function of the transit duration. In turn, the transit duration depended on the star type and planetary semi-major axis. At the time of the Mission development, only measurements of the Sun provided a sufficiently precise power spectrum that could be used as a model for other stars. Its power spectrum was known at a precision of approximately 10 ppm for periods appropriate for transit durations (i.e., several hours). The power spectrum[17] used in the model was based on the data[18] from the Differential Absolute Radiometer (DIARAD) instrument aboard the Solar and Heliospheric Observatory (SOHO) spacecraft appropriate for an average activity level. In the models, the SNR was calculated from ratio of the integral of the frequency response of a transit of a given duration and depth to the square root of the integral of the power spectrum of the expected noise: a combination of the non-white solar noise to which white noise associated with instrument and shot noise were added[19]. The SNR was then scaled by the square root of the number of transits, as determined from the orbital period.

Models of the exoplanet characteristics and distribution

Two complementary models of planetary systems were used for the calculations. The first model (Model 1) provided the science value for small planets in the HZ. This model assumed that there was only one planet in each planetary system and that it was in the HZ. However, for this model, the size of this planet was varied to estimate the mission capability for assumptions as to planet size. In particular, the expected number of detectable planets was tabulated for sizes of 0.53, 1.0, 1.3, 2.0 $R_\oplus$; i.e., for Mars-, Earth-, and two super-Earth-sizes. The approximate value of the semi-major axis and the period appropriate for the HZ for each star type was computed from the results in Ref. 20.

The second model (Model 2) considered only non-HZ planets and assumed that the size of each planet to be 1.0 $R_\oplus$. The semi-major axes of these planets were set to 0.05, 0.1, 0.2, 0.4, 0.8, 0.9, 1.0, 1.2, and 1.5 AU to explore the region near the HZ of a variety of stellar types. According to a study[8], a peak near 1 AU should occur in the occurrence frequency of Earth-size planets for stellar masses between 0.5 and 1.5 times the mass of the Sun ($M_\odot$). To avoid double counting planets in the HZ, the results for a planet that overlapped the HZ position of a specific star-type were omitted for the calculations used in Model 2. Although a distribution of planets as closely spaced as specified here might result in a dynamically unstable situation for some planets, the objective of the MF was to explore the statistical distribution of planets versus semi-major axis.

To better represent the Solar System, the model assumed that only 3 planets (representing Venus, Earth and Mars) exist in this region. Therefore the results from Model 2 were multiplied by 2/9. This choice meant that the combination of Models 1 and 2 considered three planets in and near the HZ.

Assignment of Values for Science Products

For the model calculations, planets with semi-major axes larger than 1.5 AU were ignored because planets in such orbits wouldn't provide the minimum number of transits (i.e., 3) during the Mission life time of 4 years and because most such orbits would not be in the HZ for solar-like stars. Planets larger than twice the size of the Earth were not considered because they would be so large and/or massive that they could be detected by ground-based observatories. Based on the calculations in Ref. 21, Earth-size planets would be distributed at random positions; i.e., Earth- and super-Earth-size planets might be expected anywhere from the orbit of Mercury to the orbit of Mars.

The overall science value was very sensitive to the choice of the values assigned to various model outcomes. For example, if the value assigned to the detection of large-size planet in short-period orbits was set equal to that of a small Earth-size planet in the HZ, then maximization of the MF would bias the design to observe many different regions of the sky for a few months each. This result occurs because of the high probability of detecting large planets in short period-orbits versus the low probability of detecting the low-amplitude signals from the transits of small planets in long-period HZ-orbits that provide few transits. Thus the choice to detect the maximum number of planets would emphasize the detection of thousands of large planets in short period orbits, but would result in the complete loss of all Earth-size planets in the HZ of solar-like stars. Instead, the model factors that represent science values in the model were weighted to emphasize the detection of small planets in the HZ. The selection of the weights for the science products was made by the PI in consultation with the mission science team and the science community.

The values generated by the MF were based on the planet size and orbital position relative to the HZ. In particular, planets detected in the HZ were considered to be more valuable than those not in the HZ (non-HZ planets). The ability to detect smaller planets was considered to be much more valuable than the capability to finding larger planets because a positive result for a small size ensures that all larger sizes would also be detected. Therefore the capability to detect Mars-size, Earth-size, 1.3 $R_\oplus$ ,and 2 $R_\oplus$, planets was assigned 40, 20, 5, and 1 points, respectively. Because ground-based observations can detect planets larger than twice the diameter of the Earth, planets larger than 2 $R_\oplus$ were not considered. Such planets were also expected to have extensive hydrogen atmospheres i.e., non-terrestrial planets incapable of supporting carbon-based life.

A value of "100 points" was assigned to the results expected from the instrument design as proposed in the Year-2000 proposal. Sixty-five percent of the total score was allotted to planets detectable in the HZ, thirty-five percent to non-HZ planets. Modifications to the Mission design or revisions to the mission assumptions were ranked relative to this "CSR performance equals100-points" case ("Case#0").

For Model 1 (only HZ-planets), the algorithm calculated: 1) the probability of orbital alignment for each star, 2) the number of transits that occur for a planet in the HZ during duration of the mission, and 3) the SNR based on the size and brightness of the star, the transit duration, quadrature sum of noise sources, and the number of transits. Then the probability of recognition was calculated from the value of the SNR relative to the threshold SNR. The probability of recognition was based on a Gaussian distribution where the difference between the calculated SNR and the threshold SNR was taken to be the significance of the detection: i.e., for a threshold SNR=7$\sigma$, and for calculated SNRs of 7$\sigma$ and 8$\sigma$, the recognition rates would be 50% and 92%, respectively. A score was assigned based on the size of the smallest planet that can be detected in the HZ for each star. This process was repeated for all of the target stars in the FOV and then the score was summed and normalized. To avoid over-counting the score for stars for which smaller planets could be detected, the score for each selected size was calculated only for the stars that were in addition to those already accounted; i.e., those with smaller-size detectable planets.

The evaluation of Model 2 (non-HZ planets) included the weighting of the results based on the rapidly decreasing geometrical probability of the orbital alignment with increasing semi-major axis. Planets very close to their star have a high probability of geometrical alignment (i.e., probability = r*/$R_{orbit}$, where r* is the radius of the star and $R_{orbit}$ is the orbital radius). For example, planets at 0.05 AU from their star are 20 times more likely to be correctly aligned than those at 1AU. Furthermore, the number of transits observed during the duration of the mission is much larger for planets with small semi-major axes than for planets with large semi-major axes. (*Kepler*'s third law; i.e., the period squared is proportional to semi-major axis cubed.) Thus equal-size planets in the line-of-sight with a semi-major axis of 0.05 AU present about 89 times the number of transits of a planet with a 1 AU orbit. This results in the shorter-period planets having a total SNR approximately 9 times higher than long-period planets and are therefore being much more easily recognized. Based on these two factors, most of the detections (and therefore most of the MF score) would be from planets orbiting too close to their star to be in or near the HZ.

However, planets near the HZ were considered to have a higher science value because such detections were more consistent with the goals of the Mission. To reduce the biases introduced by small semi-major axes, the MF value of each recognized planet in Model 2 was weighted by multiplying its value by the square of its semi-major axis (in AU). The combination of weighting of the score for the position of each planet and limiting the contribution to the total score of 35% for non-HZ planets ensured that the MF scores would be dominated by the mission performance with respect to small planets in and near the 1 AU.

A final score was computed by adding the values for Models 1 and 2. It is recognized that different weights and normalization factors could have been chosen for the calculations. The values chosen here were considered appropriate by the PI and his science team. An outline of the MF approach is presented in Appendix 1.

The orbital period and transit duration were then calculated for the selected values of the mass of each star type, the selected semi-major axis, and the semi-major axis for the HZ. The number of transits observed during the mission was calculated from the orbital period and mission duration. The calculated value of the transit duration was used to estimate the contribution of stellar variability noise and shot noise associated with the integrated photon flux. The value of stellar variability contribution to the total noise was derived from the power spectrum of the Sun for the calculated transit duration. Because the solar noise increases with period, the SNR of the transits depends on the transit duration[19]. A value for the instrument noise was based on "shot noise" and the effect of "jitter" associated with tracking error.

Once the properties of the transit and the number of transits that occur during the mission were computed, the noise level and SNR were computed for each star. Although not discussed here, calculations then reduced the SNR to account for the expected number of missed transits and missed data that typically occur in a mission.

In the next section, the Mission expectations are presented for the "point-design"; i.e., the expected results for the values of the Mission parameters specified in the CSR. Section 5 will consider the effects of varying values of the mission parameters.

## 4. Expected results from mission operations

<u>Mission Capability with Respect to Planet Size and Spectral Type</u>
The expected number of detectable planets increases with the assumed size of the planets until all the planets in the FOV are sufficiently large with respect to their parent star to produce transit SNRs well above the detection threshold.

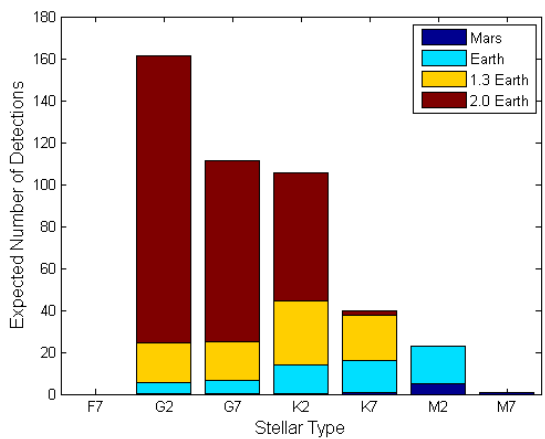

**Figure 3. Expected number of detections of planets in the HZ versus spectral type.** The coloring of the bars is related to the size of the planets found at each star type.

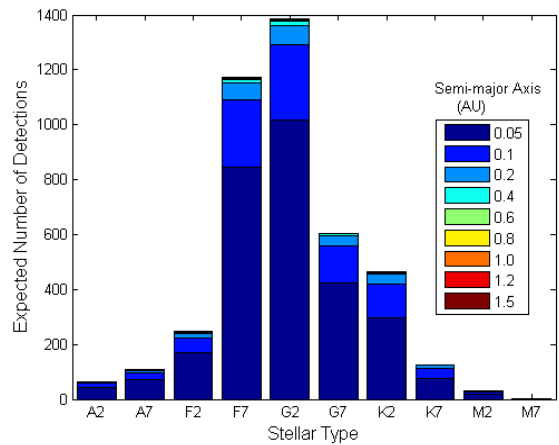

**Figure 4. Expected number of detections of planets not in the HZ versus spectral type.** The coloring of the bars is related to the size of the semi-major axis for a given star type.

Figure 3 illustrates the distribution of the expected number of detectable HZ planets with planet size and stellar type assuming that all the mission assumptions for Case #1 are appropriate. The colored bars represent expectations for **four** different assumptions; 1) all stars have only Mars-size planets, 2) all stars have only Earth-size planets, 3) all stars have only 1.3 $R_\oplus$ size planets, and 4) all stars have only 2 $R_\oplus$ planets. In particular, for the assumption that all stars have HZ planets that are similar in size to Mars, then the results shown by the dark blue bars are applicable. Figure 3 shows that the Mission has the capability of finding Mars-size planets mostly for M-type stars. However, if larger planets are common, then many more planets orbiting a wider range of stellar types are expected as shown by the sum of the lighter-colored bars. The predicted numbers of detectable planets for each stellar type and for any of the chosen planets sizes (i.e., Mars-size, Earth-size, 1.3 $R_\oplus$, and 2 $R_\oplus$) are indicated by the sum of the heights of the appropriate bars; e.g., the detection of 40 planets with Rp ~ 1.3 $R_\oplus$ is expected for stars of spectral type K2. Note that for the M-type stars, the bars are saturated; considering a larger planet size does not lead to more detections because the amplitudes of the transits of small planets are already larger than the detection threshold and because there are no more M-type stars in the FOV that have magnitudes less than the limiting value.

Summing the values of the blue- and aqua-colored bars indicates that 66 Earth-size planets could be detected in the HZ if all stars had such planets. If only 10% of the stars have Earth-size planets, a statistically useful number could still be expected. If a null result was found instead, the Mission results would indicate that such planets were rare; i.e., substantially less than 10% of stars had such planets in their HZ. The expected number of detectable planets which have a radius ~1.3 $R_\oplus$, is more than double that for Earth-size planets. Note that planets with Rp ≤1.3 $R_\oplus$ are expected to have a rocky composition[22]. Consequently, it is likely there would be a sufficient number in each of the spectral types G2 through M2 to provide a coarse estimate of the occurrence frequency of small/rocky planets vs spectral type. If most stars had planets twice Earth-size, then 520 detections would be expected for planets in the HZ. Thus the Mission capability should be sufficient to address the Mission goals for the planetary occurrence frequency for small planets, their occurrence frequency versus spectral type, and their size distribution.

Figure 4 shows the results for the expected occurrence frequency of planets that are not in the HZ (i.e., non-HZ planets). The calculations assume that at 8 non-overlapped values of the 9 values of the semi-major axes (specified in the figure legend) there is the possibility of an Earth-size planet. The model assumes that each planetary system will have an average of 3 planets; one in the HZ and two in non-HZ orbits.

It is clear from Figure 4 that most of the detections are expected at the smallest values of the semi-major axes (i.e., at the shortest orbital periods). This result is expected because the probability of transits is proportional to the ratio of the stellar size to the orbital size and because of the increased value of the SNR caused by the large number of transits for planets in short-period orbits. Summing the values in Figure 4 indicates that over 4200 non-HZ planets should be found provided that planets at least as large as the Earth are common. Although this value might appear to be surprisingly large, it is actually similar to the values[23,24] found for the *Kepler* Mission and is expected for the hypothesis that most of the 170,000 stars have planets.

Note that the predicted Mission capability for non-HZ planets would increase if larger than Earth-size planets were considered. The choice that all non-HZ planets be Earth-size was made to insure that the Mission design emphasized Earth-size planets in and near the HZ as stated in the Science Goals (Section 2).

In contrast to the results for HZ-planets shown in Figure 3, Figure 4 indicates that the Mission has the capability of detecting many non-HZ planets orbiting A- and F-type stars. This is the result of choosing the target stars based on brightness without regard for spectral type (i.e., Cases #0 & #1) and that HZ-planets orbiting such stars have orbital periods too long to show a minimum of three transits during a 4-year mission duration.

## 5. Analysis of some factors controlling science results

<u>Number and selection of target stars</u>

The number of expected detections as well as the accuracy of the estimates of the occurrence frequency and the distributions of planets with size and stellar spectral type depend on the number of useful stars that can be monitored with the required precision. Monitoring a large number of stars was required to get statistically meaningful results because the geometrical probability that a planet in a particular orbit will show transits is very low[25]. The probability is the diameter of the star divided by the diameter of the orbit: i.e., about 0.1 for planets with orbital periods of a few days and 0.005 for planets like that of the Earth orbiting in the HZ of a star like the Sun. Assuming that every star has a planet large enough to be detected and that the planet is in a properly aligned orbit in the HZ of a solar-like star, then several hundred planets should be detected based on monitoring 100,000 stars. During the development of the CSR, it was recognized that an increase in the instrument memory capability would allow the number to be increased to 170,000 stars. However, increasing the number of stars implies a reduction in "time-on-target" due to a longer period[2] required to orient the spacecraft toward the Earth to download the data.

Maximizing the number of detected planets requires maximizing the number of useful stars in a single FOV. Although the number of stars in a fixed FOV grows almost exponentially with the acceptable upper limit to the magnitude, only a fraction of the added stars are suitable for the search because some of these are too dim or too large to provide a SNR above the threshold value.

Prior to the approval of the Mission for development, the choice of target stars for the *Kepler* Mission was necessarily based only on apparent brightness because stellar sizes were unknown for most stars dimmer than 9[th] magnitude. However, after the start of Mission development, members of the *Kepler* team made multi-band color measurements of 2 million of the stars in the selected FOV to determine their spectral type and thereby, their size[26]. Consequently it was possible to obtain a selection of target stars that maximized the detectability of small planets; i.e., to do "cherry-picking".

Cherry-picking consists of selecting those stars with the size and apparent brightness to provide the highest probability of showing a pattern of transits that are above the detection threshold for Earth-size planets in the HZ. Thus stars with the lowest noise are not necessarily the best choice. For example, although two stars of equal apparent brightness but of dissimilar size have similar levels of shot noise, the smaller star (i.e., later spectral type and lower mass) will show more transits for planets in the HZ during the duration of the mission; and these transits will have larger amplitudes than those for the larger star. The probability of the geometrical alignment of the orbit with the line-of-sight and the duration of the transits also enter into the selection of the "cherry-picked" targets.

An algorithm[27] was written to choose the most appropriate stars as a function of the number of stars monitored in a fixed FOV. (The optimized selection of stars for the actual Mission is discussed in Ref. 28). In this study, the following factors were considered:

---

[2] Downlinks were performed once per month and took ~12 hours to complete.

1) stars considered that are too dim to give a detectable signal for an Earth-size planet can still produce detectable signals for transits of the larger planets,
2) for a given planet size, a small star will produce a larger relative signal than a larger star and thus show detectable transit amplitudes even though it is dimmer than a somewhat larger and brighter star,
3) the probability of an orbital alignment is proportional to the size of a star, and
4) the orbital periods for planets in the HZ of small (low-mass) stars are short and therefore provide many transits during the mission duration thereby increasing the SNR compared to the longer periods for HZ planets of larger, more-massive stars.

The MF was run to estimate the expected results of observing for three selections of stars in a fixed FOV near the galactic plane. Table 1 shows results for the three cases: Case #0: the selection of the 100,000 brightest stars between 9th and 15th magnitude as specified in the original proposal; Case #1: 170,000 brightest stars between 9th and 15.5 magnitude; and Case #2: 170,000 "cherry-picked" stars between 9th magnitude and 16.5 magnitude that were chosen to maximize detectability of small planets near the HZ.

For Cases #0 and #1, it was assumed that all stellar types are observed because the information on stellar type and size needed to choose the most appropriate stars was not available. For Case #2, it was assumed that a ground-based survey to classify the stars was conducted prior to the launch of the Mission. Thus, spectral types earlier than F5 were rejected and replaced by later types. An examination of Table 1 shows that the MF value increased by 20% when the number of stars monitored are increased from 100,000 to 170,000 while the number of small planets ($R_p \leq 1.3\ R_\oplus$) in the HZ and planets not-in-the HZ increase by 25% and 31%, respectively. However when cherry-picked target stars are chosen rather than the 170,000 brightest stars, the MF for small planets in the HZ increases by 40% but the number of non-HZ planets decreases by 12%. The decrease in the number of non-HZ planets occurs because Case#2 selects for slightly dimmer and smaller stars that can provide detections of $2R_\oplus$ planets in the HZ. This choice decreases the number of short-period planets that would have been detected around the brighter and larger stars. The Mission capability to detect $2\ R_\oplus$ planets in the HZ increases by 280% from Case #0 to Case #2.

| Table 1. MF Value, Number of Detectable Planets in the HZ, and Number of Detectable Planets not in the HZ | | | | | | |
|---|---|---|---|---|---|---|
| Case | Maximum magnitude limit | Number of stars monitored | Calculated Merit Function Value | Number of detectable HZ planets ($R_p \leq 1.3 R_\oplus$) | Number of ~$2R_\oplus$ planets in HZ | Number of detectable non-HZ planets |
| 0 | 15.0 | 100,000 brightest | 83 | 125 | 278 | 3205 |
| 1 | 15.5 | 170,000 brightest | 100 | 156 | 442 | 4200 |
| 2 | 16.5 | 170,000 cherry-picked | 140 | 255 | 788 | 3707 |

In Table 1, the number of 2 $R_\oplus$ planets are shown separately from the planets with radii ≤1.3 $R_\oplus$ because the larger planets might be too large to have a rocky composition[22].

Table 1 can be used to examine the effects of changing the assumptions that scale linearly with the number of stars; e.g., the planetary occurrence rate and/or decreasing the size of the FOV. However, many of the expected mission results have non-linear dependencies on the model assumptions. Therefore model calculations are required to estimate their effects. Figures 3 and 4 display the results of some of these calculations.

Comparison of the numbers of detectable planets in columns 5 and 6 of the Table indicates that most of the increase in the number of HZ planets results from the inclusion of the "large" 2 $R_\oplus$ planets. These planets have 4 times the signal amplitude of the 1 $R_\oplus$ planets and therefore are much easier to detect among the many dim stars in the FOV. The capability of the Mission to find several hundred planets in the HZ reduces the risk that the Mission might not find any planets in HZ if the occurrence frequency were as low as 1%.

The model results shown in Figure 4, indicate that the number of detectable non-HZ Earth-size planets is much larger than those found in the HZ (Note the greatly increased vertical scale). Again it is clear that a useful distribution of occurrence frequency vs semi-major axis and stellar type can be expected even if the frequency of small planets is 1% or less. Most of these are in short period orbits that produce many transits, have a high probability of orbital alignment, and thereby have a high probability of detection. These calculations assume that all non-HZ planets are Earth-size. Clearly many additional planets would be detectable if the assumption of larger size planets were allowed. The choice of only 1$R_\oplus$ planet-size for non-HZ planets assures that the Mission design emphasizes Earth-size planets in and near the HZ as stated in the Science Goals (Section 1). Although the expected number of non-HZ planet detections is quite large, the value is similar to that detected by the *Kepler* Mission; i.e., a total 4600 confirmed planets and candidates[24]. A comparison of the number of non-HZ planets presented in Table 1 shows that the selection of the dimmer, cherry-picked stars results in the detection of a slightly lower number of non-HZ planets.

Figures 5 and 6 display the results of model calculations for the value of the MF and numbers of exoplanets versus number of stars observed and for different selections of target stars.

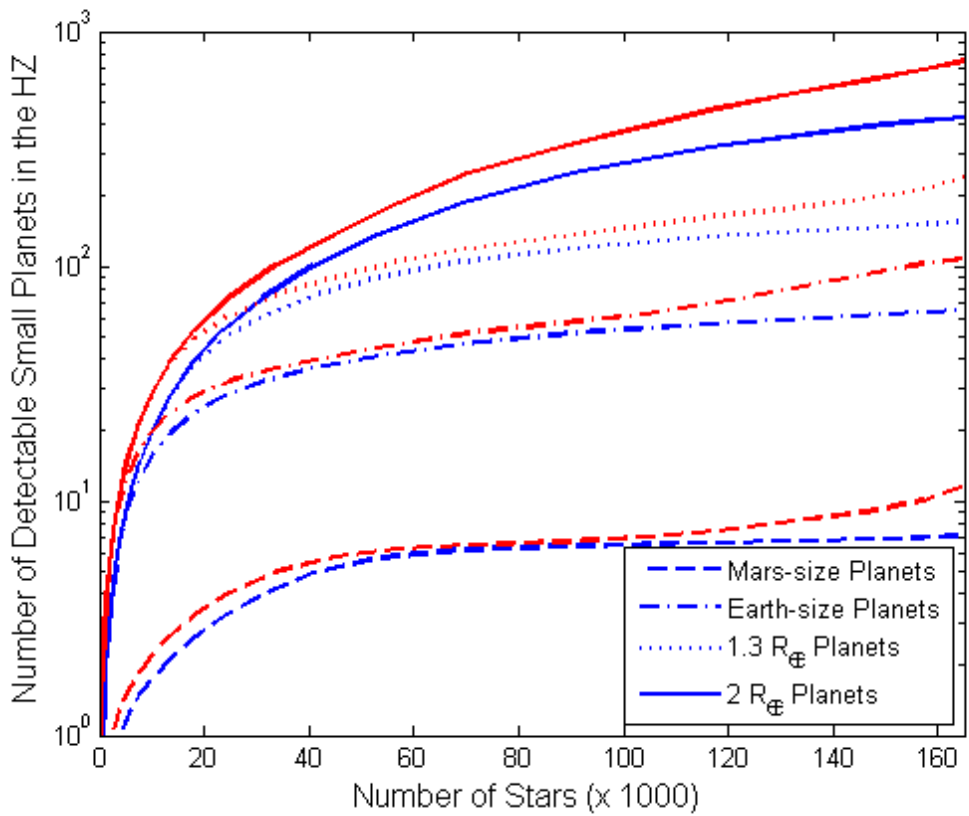

**Figure 5. Mission capability of detecting planets of various sizes in the HZ versus the number of stars observed for a fixed FOV.** The blue lines of each pair show the results for the choice of the 170,000 brightest stars while red upper line of each pair shows the results when 170,000 "cherry-picked" stars are chosen. The red line lies above the blue line because all early-spectral-type stars (i.e., large stars) have been removed from the cherry-picked selection.

The pair of dashed lines at the bottom of Figure 5 imply that there are very few stars that are small- and bright-enough to provide a SNR sufficiently large to detect Mars-size planets in the HZ. However when we consider Earth-size and larger planets, then much larger fractions of the stars show detectable transits because of the increased SNR. For these planets, increasing the number of observed stars significantly increases the number of expected detections regardless of the fact that many of the additional stars will necessarily be dimmer. The pair of solid lines in Figure 5 shows that the Mission capability to detect large terrestrial-size planets continues to increase well past the value of 100,000 stars that was assumed for the point design presented in the original proposal. The red curves show that a further gain in the number of HZ planets can be obtained by "cherry-picking", i.e., determining the spectral type and luminosity class of each star in the FOV and then choosing only those that provide largest transit signals. Much of the gain shown by the red curves relative to the blue curves is due to the replacement of early-type stars with later spectral types.

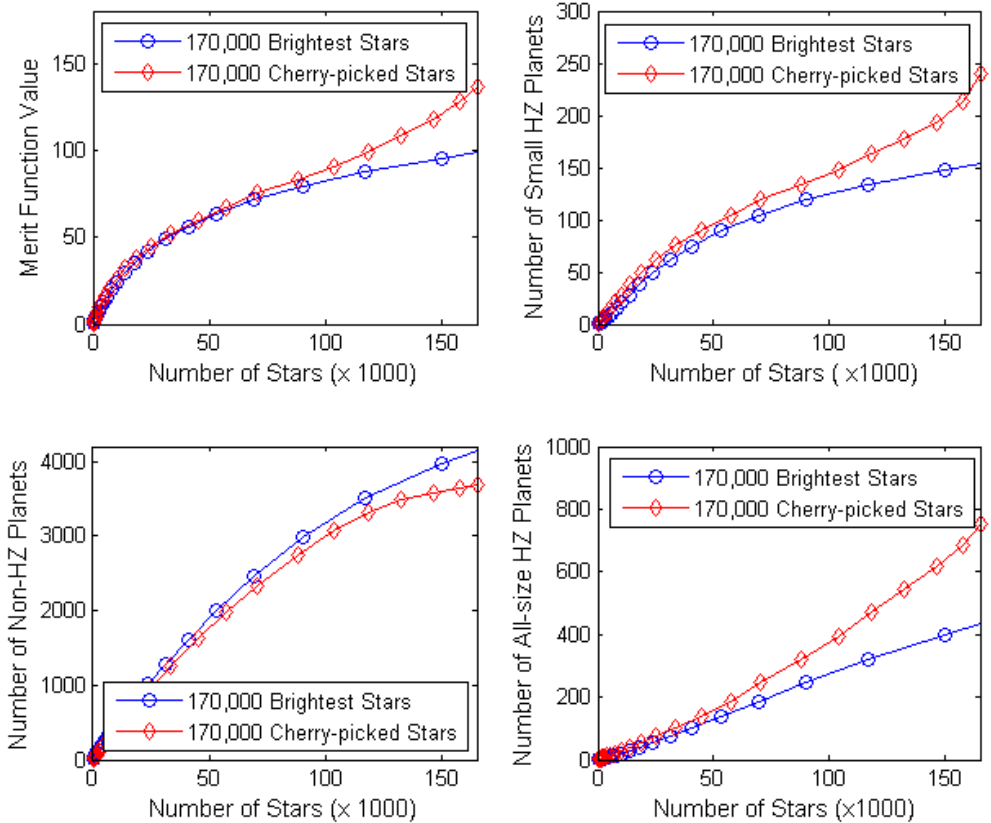

**Figure 6.** Model results for the values of the MF, number of small planets (Rp ≤ 1.3 R_⊕) in the HZ, total number of planets (Rp ≤ 2 R_⊕) in the HZ, and the number of non-HZ planets versus the number of monitored stars. The results shown in Figure 5 and 6 are based on a statistical table[16] for the number of stars versus size and magnitude.

Computing values for the MF for HZ planets at a particular value for the number of stars monitored consists of evaluating the number of stars that could show detections of the smallest planet size, multiplying that number by the science value for the appropriate planet size, and then summing the product of the additional number of planets at each size and value;  MF= Σ(n_i−n_{i-1})* v_i , i=1 to 4  and where n_i is the number of detectable planets of each of the four planets sizes and v_i is the value of the detection for planets of that size. (where n_0 =0.)

The curves in upper left-hand panel of Figure 6 represent the calculated MF for three cases: Case#0; monitor 100,000 brightest stars in the FOV, Case#1: monitor 170,000 brightest stars in the FOV, and Case#2: monitor 170,000 stars chosen that have the highest probability of showing detectable transits. (Note: To avoid saturating the detectors, only stars 9th magnitude and fainter were considered in all calculations.) Both the red and blue curves indicate that there is a substantial increase in the science product for increasing the number of monitored stars from the 100,000 in the original proposal to 170,000 in the CSR. The red curve indicates that a pre-launch program to classify all the stars in the FOV provides a substantial increase in the MF when the results are used to select the most promising stars.

Comparison of the curves showing the numbers of detectable planets in the upper- and lower- right-hand panels of Figure 6 indicates that most of the increase (note the changed range of ordinate values) in the number of HZ planets results from the inclusion of the "large" 2 $R_\oplus$ planets. These planets have 4 times the signal amplitude of the 1 $R_\oplus$ planets and therefore are much easier to detect among the many dim stars in the FOV. The capability of the Mission to find several hundred planets in the HZ reduces the risk that the Mission might not find any small planets in the HZ if the occurrence frequency were as low as 1%.

In the lower left-hand panel of Figure 6, the model results show that the number of detectable non-HZ Earth-size planets is much larger than those found in the HZ (Note the greatly increased vertical scale). It is clear that a useful distribution of occurrence frequency vs semi-major axis and stellar type can be expected even if the frequency of small planets is 1% or less. Most of these are in short period orbits that produce many transits, have a high probability of orbital alignment, and thereby have a high probability of detection. These calculations assume that all non-HZ planets are Earth-size. Clearly many additional planets would be detectable if the assumption of larger size planets had been used. The choice of only 1$R_\oplus$ planet-size for non-HZ planets assures that the Mission design emphasizes Earth-size planets in and near the HZ as stated in the Science Goals (Section 1). Although the expected number of non-HZ planet detections is quite large, the value is similar to that detected by the *Kepler* Mission; i.e., a total 4600 confirmed planets and candidates (Ref. 24). The curves in the lower left panel also show that choosing the dimmer cherry-picked stars will result in the detection of a slightly lower number of non-HZ planets.

Effects of Star Selection Criterion on Mission Capability
The left-hand and right-hand panels of Figure 7 compare the mission capabilities for two selections of target stars. The left-hand panel displays the results for the selection of 170,000 brightest (Case#1) stars including all spectral types while the right-hand panel is for a selection of 170,000 "cherry picked" stars (Case#2) that did not include early-type stars. Selection of Case#2 provides a dramatic increase in the Mission capability to detect HZ planets for later spectral-type stars with a modest reduction of detections for G2-type stars. Choosing cherry-picked stars increased Mission capability to detect both small planets and 2$R_\oplus$ planets by 78%. A comparison of the results in the left- and right-hand panels in Figure 7 shows that a more uniform sampling of the number of planets versus spectral type is obtained when cherry-picked stars are selected. In fact, the *Kepler* Mission did include a pre-launch activity[26] to classify the spectral type of stars in the FOV to provide the spectral classifications needed for Case#2.

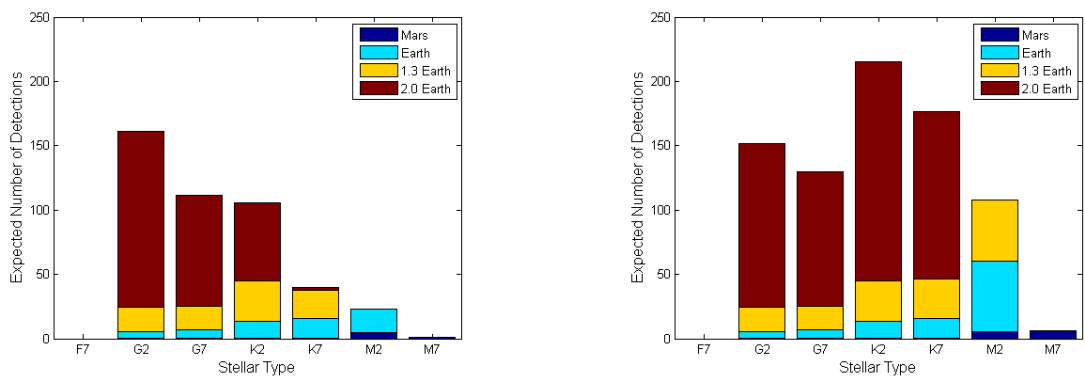

**Figure 7.** Expected number of detections based on the instrument capability, characterization of the stars in the *Kepler* FOV, and the assumptions of one terrestrial planet in each HZ and a requirement of detecting a minimum of 3 transits. Results in the left-hand panel are for the selection of 170,000 brightest stars (Case#1). Right-hand panel shows results for 170,000 "cherry-picked" stars (Case#2). The predicted numbers of detectable planets for each stellar type and for any of the chosen planets sizes (i.e., Mars-size, Earth-size, 1.3 Re, and 2 Re) are indicated by the sum of the heights of the appropriate bars.

Both panels in Figure 7 indicate that no planets are detectable orbiting F7 stars. This result stems from the requirement that at least 4 transits be observed during a four year mission and the fact that planets in the HZ of F7 and earlier (i.e., hotter) types have orbital periods too long to provide even 3 transits during the 4-year mission duration.

In contrast to what is expected for HZ planets, Figure 8 indicates that the detection of non-HZ planets orbiting stars hotter than G2-types can be expected. This occurs because planets in inner orbits will show a sufficient number of transits to satisfy the criteria of a minimum of 4 transits and provide a SNR greater than the detection threshold. It is clear from the length of the dark-colored segments as a fraction of the total height of each bar that the results are strongly biased toward the detection of innermost planets.

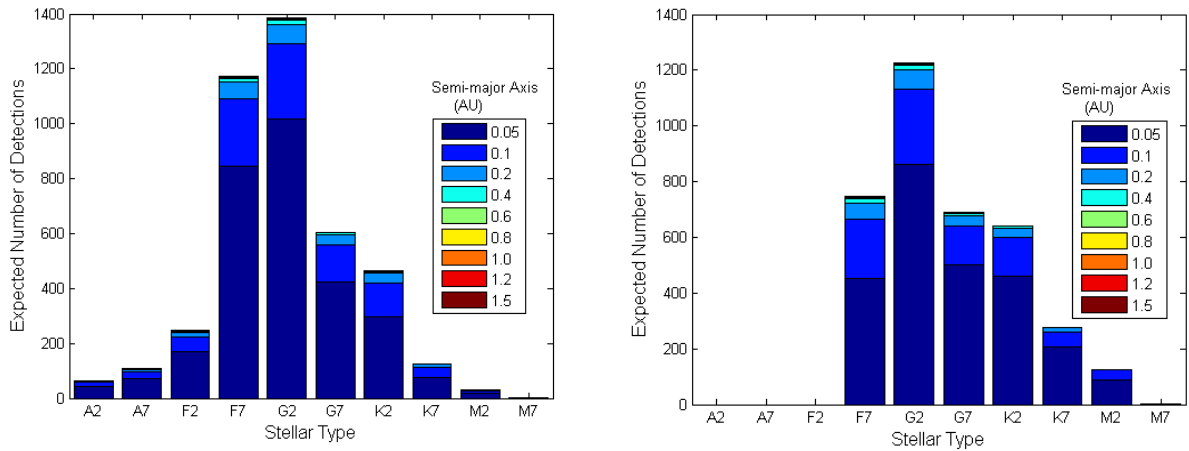

**Figure 8.** Expected number of detections for non-HZ planets distributed over a range of semi-major axes for a mission duration of 4 years. Note the greatly enlarged values on the ordinate compared to those in Figure 7. The coloring of the bars is related to the number of planets found at each semi-major axis for a given star type. Left-hand panel for Case #1: 170,000 stars selected based on their brightness. Right-hand panel shows the expected detections for Case #2: stars chosen based on the expected SNR of small-planet transits.

A comparison of the two panels in Figure 8 illustrates the effects of target star selection on the mission capability for defining the distribution of non-HZ planets. Similar to the results for HZ planets, smaller numbers of planets orbiting F7- and G2-type stars are expected for Case#2 while the number of planet detections for later-type stars is enhanced.

The results shown in Figures 7 and 8 indicate that the Mission capability should be sufficient to detect several hundred terrestrial planets in the HZ and several thousand in the non-HZ regions if they are common. Even if only a fraction of the numbers calculated here were found, useful estimates of the occurrence frequency and distributions with planet size and semi-major axis would be obtained. Furthermore, if only null results were obtained, the results would still provide statistically meaningful upper bounds and would imply that Earth-sized planets in our galaxy are rare.

Although the discussion above demonstrates that the expected number of detections is large, the predictions depend not only on prevalence and characteristics of planetary systems, but on the actual performance of the instrument. Examples of some of the risks associated with instrument and mission characteristics are given next.

## 6. Results from several risk studies

Effects of reduced mission duration
Reduction of the mission duration could result from major component failure or insufficient funding. Figure 9 shows the effects of reducing mission duration. A comparison of the upper panels shows that the reduction of the mission duration from 4 to 3 years would result in the loss of mission capability to detect Earth-size planets in the HZ of G2-dwarfs because too few transits would be observed. Because a major goal of the Mission is the determination of small planets in the HZ of Sun-like stars (i.e., G-type stars), the loss of G2-type stars would cause a serious loss of science merit. In particular, the MF value would fall from 120 to 84 for Case#1 and from 169 to 123 for Case#2; i.e., the science value would be reduced by a third. Figure 9 shows that a two-year mission would be too short to detect planets orbiting any G dwarfs  while a 1-year mission could be expected to be sensitive only to planets orbiting M-dwarfs.

Nevertheless, a mission duration of only 1 year would still be expected to produce nearly 100 detections of HZ planets. Thus a null result from a severely curtailed mission duration would still be meaningful.

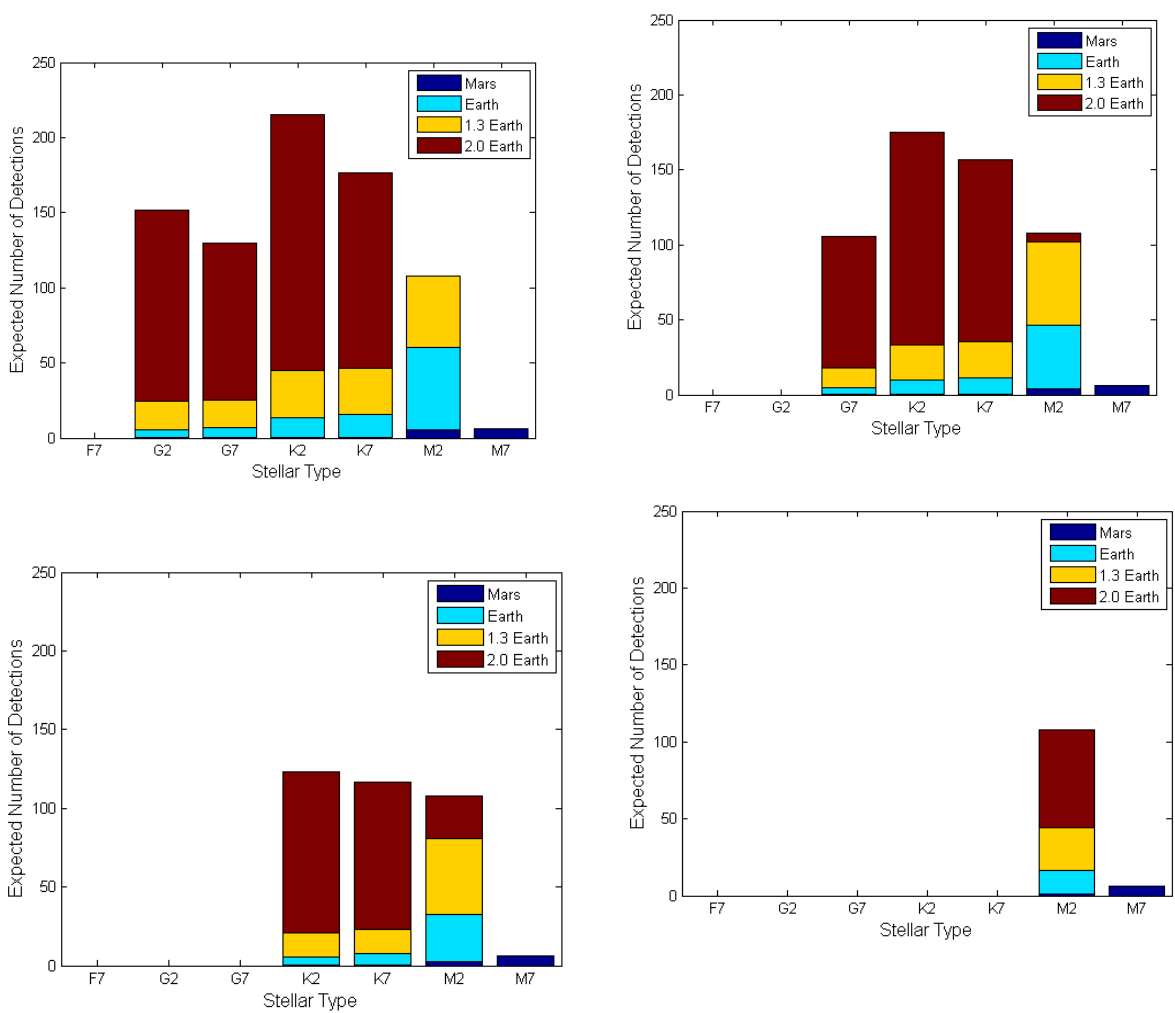

**Figure 9.** Mission capability to detect HZ-planets when the mission duration is reduced from 4 years to 1 year for cherry-picked stars (Case#2). (Left to right and then down; 4, 3, 2, & 1 year mission durations.)

Effects of higher than expected instrument noise

As the value of the instrument noise increases, not only does the number of expected planetary detections decrease, but the relative occurrence frequency with respect to the stellar type changes; i.e., another bias is introduced. Figure 10 presents model results for G, K, and M spectral types for values of the instrument noise level relative to the point design. Two sets of curves are shown. The curves shown in black represent Case#1 and curves shown in red represent Case #2. The vertical axis displays the fraction of all possible planet detections (i.e., the exoplanet population) for each star type.

The black curve marked by square symbols (Case #1) shows that the Mission has the capability of detecting all the small planets orbiting M-dwarfs until the instrument noise exceeds three times the point design value. However, the curve marked in red for Case #2 indicates that the detected fraction of the total population decreases rapidly as the noise level exceeds the value for the point design. The difference in the results for Cases #1 and #2 is caused by the selection of M-dwarfs as faint as 16.5 magnitude to maximize the number observed (~ 9400) for Case #2 while observing only the ~ 2000 M-dwarfs that are brighter than 15.5 magnitude for Case #1. As the instrument noise increases, it readily overwhelms the many faint stars associated with Case #2. Nevertheless, an examination of an earlier figure (i.e., Figure 7) shows that the number of small planets detectable in the HZ of M-dwarfs is several times larger for Case# 2 compared to that for Case #1.

The situation shown for K-dwarfs in Figure 10 is analogous to that for the M-dwarfs in that a larger fraction of the population of planets orbiting K-dwarfs is detectable because Case #1 selects only the brightest stars. Because Case #2 selects for K-dwarfs as faint as 16th magnitude, the number of selected stars is larger (68,000 versus 23,000), but many of the planets transiting fainter stars for the Case # 2 selection are not detectable at increased noise levels. The tripling of the detectable fraction for K-dwarfs shown in Figure 7 is mostly due to the removal of A- and F-type dwarfs and their replacement by K-dwarfs when Case#2 is selected.

The decrease in the fraction of the population of G-dwarfs with increasing instrument noise is nearly independent of the Case number.

It should also be noted that changes in the instrument noise level cause changes in the relative number of planets with spectral type as seen in both Figures 7 and 10.

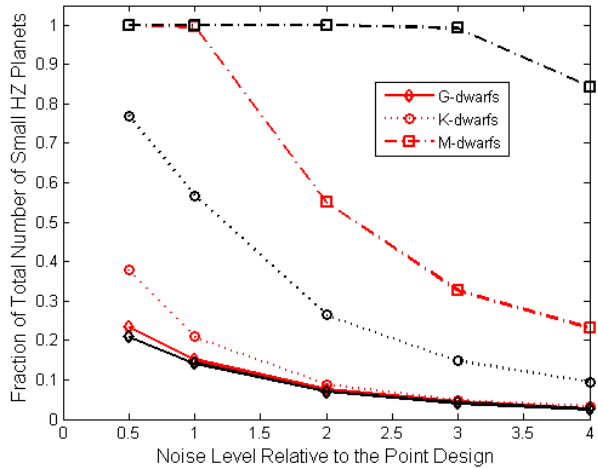

**Figure 10.** Fraction of the population of small planets (Rp ≤1.3 R⊕) in the HZ that can be detected in the FOV. Black curves represent Case#1 while red curves represent Case#2.

<u>Effects of raising the detection threshold</u>
During the Mission design, investigations[12] using the MF were made regarding various system and stellar noise characteristics. However, the effects of non-white noise and astrophysical false-positive events were difficult to predict. The *Kepler* Mission design assumed that a threshold of 7σ for detection would be satisfactory to reduce the number of statistical false alarms to ≤ 1. However this value was not sufficient to avoid frequent false-positive (i.e., astrophysical events that mimicked a planetary transit) and false-alarm events (i.e., statistical noise). False-positive events[29] were often caused by images from eclipsing binary stars in or near the target star images. Both types of events often led to frequent false-alarms that wasted valuable resources needed to confirm a planet.

Consequently, rather than following up all 7 σ detections, often only those events with much higher values attracted the resources needed for follow up observations. Figure 11 displays the predicted effects on the detection rates as the required threshold level is increased from 7σ to 12σ. The results imply that many more planets remain to be detected if the false alarm and false-positive rates can be reduced sufficiently to allow 7σ detections to be confirmed.

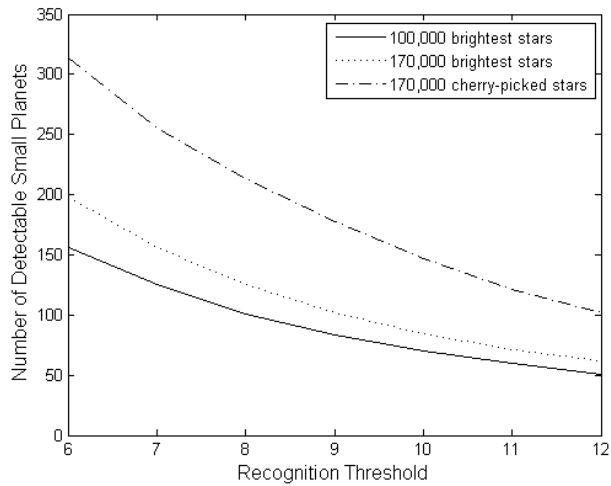

Figure 11. Predicted number of detectable small planets (Rp ≤ 1.3 R⊕) versus the value of the recognition threshold. Note that a value of 7 was chosen for Mission point design and the calculations in this paper.

## 7. Brief comparison of model predictions with Mission results

Only a rough comparison with the *Kepler* Mission results is possible because of the many differences between the model study and the actual *Kepler* survey; in particular

1) The model assumes that the planets are placed at specific locations whereas the actual number of discoveries occurred along a continuum of planet sizes and semi-major axes.
2) The model assumed that all non-HZ planets were no larger than Earth-size
3) 50% of the target stars in the *Kepler* survey were probably binary or multiple-star systems. The non-repetitive transit patterns from planets orbiting binary stars were much more difficult to detect than those from single stars[30] and because the secondary star diluted the transit depth.
4) Stellar variability[31,32] and system noise in the *Kepler* survey were much higher than expected and than was assumed here.
5) The high false-positive rate due to many faint stars in or near target apertures required ignoring transit patterns with an SNR much less than ~12.
6) The model results for small planets included many planets orbiting very faint stars that would not have been detectable by the follow up program.

It is clear from Table 1 and the Mission results[23,24] that the model predictions for the occurrence rate of HZ-planets are much higher than that found by the Mission (255 versus 23 for Rp ≤1.3 R⊕, and 778 versus 104 for Rp ≤ 2 R⊕)[3]. However, note that the Mission found an additional 184 planets larger than 2 R⊕ in the HZ. Such situations would exclude smaller planets from the HZ.   A comparison of the predicted number of small (Rp ≤ 1.3 R⊕) planets in the HZ with those actually discovered shows that the model assumption that every star has such a planet in its HZ over-predicts the observed number by a factor of 11. The prediction is a factor of 7 for the larger, more easily recognized 2 $R_{\oplus}$ planets. Assuming that half the target stars were binary and that these seldom provided detectable planets in the HZ, compensation for this factor would double the fraction of small planets detected (i.e., 23/255) to ~46/255. (The Mission detected only 10 planets orbiting binary stars compared to total of 4700 detected planets.) This result implies that the ratio of stars that have small planets in their HZ is of the order of 20%. This value is mid-range to the values near 1% (Refs. 33 & 34), ~10 to 40% (Refs. 35-37), and near 100% (Ref. 38). The wide range of estimates is due to both the uncertainty in the statistics of the observations and to the varied assumptions with respect to ranges of planet size and orbital period chosen for the computations.

---

[3] Values for the Kepler Mission results are based on the NASA Exoplanet Archive (https://exoplanetarchive.ipac.caltech.edu/docs/counts_detail.html) as of 02/19/2020. The predictions are based on Case #2.

The MF predicted number (i.e., ~3700) for non-HZ planets is similar to that found by the Mission (i.e., 4488). Recognizing that the model assumes that all stars have 2 non-HZ Earth-size planets with semi-major axes $\leq$ 2 AU, the similarity between the predicted and observed values implies that multi-planet systems are common and that many stars have at least two planets.

## 8. Summary

The MF is an algorithm designed to produce numerical values for the science return given a set of inputs representing an instrument/spacecraft point-design, estimates of the stellar structure of galaxy, exoplanet size and frequency distributions, and instrument noise. Examples of the consequences of mission risks were explored.

Several examples demonstrated the usefulness of the MF; in particular the results show that:
1) Increasing the number of stars observed from the 100,000 brightest stars in the original Mission proposal to 170,000 substantially increases the mission capability to detect both HZ- and non-HZ planets by 59% and 31%, respectively).
2) Additionally, a pre-launch stellar observation activity to select the most promising targets ("cherry-picking") further increases the number of discoveries and better defines the occurrence frequency and the size distribution of exoplanets. In particular, the Mission capability to detect small planets in the HZ of K-type dwarfs nearly triples when stars are selected based on the expected SNR. For planets not in the HZ, cherry-picking the selection of targets stars changes the observed distribution of planets versus stellar spectral type by decreasing the number of planets associated with early-type stars while increasing those orbiting later-type stars.
3) Calculations showed that the risk that is incurred for shortened mission durations does not negate the Mission capability to place a useful estimate of the occurrence frequency of Earth-size planets even for a mission duration as short as one year. However reduction of the mission duration from 4 years to shorter periods does cause a severe loss with respect to the determination of the occurrence frequency for stars most like the Sun.
4) Evaluation of the risk that a noise level twice that stated in the CSR would prevent meeting Mission goals indicated that the capability would be reduced by approximately two, but Mission capability would still provide a useful number of planetary detections.

5) Even when the false-alarm and false-positive rates become so large that the threshold for signal recognition must be raised to almost double the point design, the expected discovery rate is still sufficiently high to accomplish most Mission goals.

The MF described here represents a tool to assess the change in the science value of a mission when the assumptions about unknowns are varied. Risks associated with uncertainties in model parameters, instrument performance, and the consequences of mission trade-offs can be assessed.

The strengths of the MF approach includes the ability to explore the effects of the many unknowns in first-of-its-kind missions and to assess the effects of mission changes during development and operation. Weaknesses include omission of the effects of false-positive events, of the effects of various approaches to data analysis, and of the limitations of the ground assets needed for confirmations.

**Acknowledgements**
The author would like to thank Riley Duren (JPL system engineer for the *Kepler* Project) for his suggestion that an algorithm be developed to quantitate the tradeoff between system parameters and the expected science output. I would also like to acknowledge the help with program development from Jon Jenkins, Jeff Garside, and Santanu Das and for reviewers' suggestions from Sally Cahill and Jon Jenkins. The *Kepler* Mission was the tenth Discovery Mission. The work was supported by the NASA Headquarter Science Mission Directorate. The NASA Ames Research Center, and Jet Propulsion Laboratory managed the Mission. The Ball Aerospace and Technology Corporation built the instrument and spacecraft. Many organizations contributed to Mission success, including: the Laboratory for Space Physics, SETI Institute, Smithsonian Astrophysical Observatory, Lawrence Hall of Science, California Institute of Technology, University of California Lick and Keck Observatories, Space Telescope Science Institute, University of Texas McDonald Observatory, and Roque de los Muchachos Observatory in La Palma.

**References:**

1. Henry, G. W., Baliunas, S. L., Donahue, R. A., et al., "Properties of Sun-Like Stars with Planets: 51 Pegasi, 47 Ursae Majoris, 70 Virginis, and HD 1147621," *ApJ* **474**:503 (1997).

2. Brown, T. M. & Charbonneau, D. 2000, "The STARE Project: a Transit Search for Hot Jupiters," in *Disks, Planetesimals, and Planets*," ASP Conference Proceedings, **219**, F. Garzón, C. Eiroa, D. de Winter, and T. J. Mahoney, Eds., pp.584-589, ASP, ISBN 1-58381-051-X, (2000).

3. Charbonneau, D., Brown, T. M., Latham, D. W. et al., "Detection of Planetary Transits Across a Sun-like Star," *ApJ* **529**:L45-49 (2000).

4. Borucki, W. J., et al., "The Vulcan photometer: a dedicated photometer for extrasolar planet searches," *PASP* **113**, 459-451 (2001).

5. Brown, T. M., Charbonneau, D., Gilliland, R. et al., "HST Time-Series Photometry of the Transiting Planet of HD 209458," *ApJ* **552**:699-709 (2001).

6. Wetherill, G. W., "Occurrence of Earth-like bodies in planetary systems," *Science* **253**, 535-538 (1991).

7. Lissauer, J., "Planet formation," *Ann. Rev Astron. Astrophys.* **31**:129-174 (1993).

8. Wetherill, G. W., "The formation and habitability of extra-solar planets," *Icarus* **119**, 219-238 (1996).

9. Morbidelli, A., Lunine, J. I.; O'Brien, D. P., et al., "Building terrestrial planets," *Annual Rev. Earth & Planetary Sci.* **40**, 251-275 (2012).

10. Mayor, M. and Queloz, D, "First discovery (RV) of exoplanet around a main-sequence star," *Nature* **378**:355 (1995).

11. Marcy, G. W., and Butler, R. P., "Detection of extrasolar giant planets," *Ann. Rev. Astron. Astrophysics* **36**, 57-98 (1998).

12. Borucki, W., Koch, D., Batalha, N., et al., "*Kepler*: Search for Earth-size planets in the habitable zone", in *Transiting Planets. Proceedings IAU Symposium* No. **IAU253**, Frederic Pont, Dimitar Sasselov, & Matthews Holman, Eds., pp. 289-299, IAU (2009) [doi 10.1017/S1743921308026513]

13. Borucki, W. J., "*Kepler* Mission: Development and Overview," *Rep. Prog. Phys.* **79**, 036901 (2016).

14. Koch, D. G., Borucki, W. J., Basri, G., et al., "*Kepler Mission* Design, realized photometric performance, and early science," *ApJL* **713**:L79-L86 (2010).

15. Schmidt-Kaler, Th., "Physical Parameters of the stars," in *Astronomy and Astrophysics – stars and star clusters,* K. Schaifers, H.H.Vogt, Eds., p.1, *Landolt-Bornstein New Series* (2b), Springer-Verlag, New York (1982).

16. Robin, A.C., Reylé, C., Derrière, S. and Picaud, S., "A synthetic view on structure and evolution of the Milky Way," *Astron. Astrophys.*, **409**:523 (2003); ADS (*erratum*: *Astron. Astrophys.*, 416:157 (2004))

17. Batalha, N. M., Jenkins, J., Basri, G. S., et al., "Stellar variability and its implications for photometric planet detection with *Kepler*," in *ESA SP-485; Stellar Structure and Habitable Planet Finding, Proceedings of the First Eddington Workshop,* B. Pattrick, Ed., pp. 35-40, ESTEC, Noordwijk, The Netherlands (2002).

18. Fröhlich, C., Crommelynck, D., Wherli, C. et al., "In-flight performance of the Virgo Solar Irradiance Instruments on SOHO," *Solar Physics* **175(2)**:267-286 (1997). [doi: 10.1023/A:1004929108864].

19. Jenkins, J. M., "The impact of stellar variability on the detectability of transiting terrestrial planets," *ApJ* **575**, 493-501(2002).

20. Kasting, J. E., Whitmire, D. P., and Reynolds, R. T., "Habitable zones around main sequence stars," *Icarus* **101**:108 (1993).

21. Chambers, J. E., and Wetherill, G. W., "The Terrestrial planets: N-Body integrations of embryos in three dimensions," *Icarus* **136**, 304-327 (1998).

22. Marcy, G. W., Isaacson, H., Howard, A. W., et al., "Masses, radii, and orbits of small *Kepler* planets: The transition from gaseous to rocky planets," *ApJS* **210**:20 (2014).

23. Coughlin, J. L., Mullally, F., Thompson, S. E., et al., "Planetary Candidates observed by *Kepler*.VII. The first fully uniform catalog based on the entire 48-month data set (Q1-Q17) DR 24," *ApJS* **224**:12 (2016).

24. Thompson, S. E., Coughlin, J. L.; Hoffman, K, et al., "Planetary Candidates Observed by *Kepler*. VIII. A Fully Automated Catalog with Measured Completeness and Reliability Based on Data Release 25," *ApJS* **235**, 38T (2018).

25. Borucki, W. J., and A. L. Summers, "The photometric method of detecting other planetary systems," *Icarus* **58**, 121-134 (1984).

26. Brown, T. M. et al., "*Kepler* Input Catalog: Photometric calibration and stellar classification," *AJ* **142**,112 (2011).

27. Borucki, W. J., Koch, D., Basri, G., et al., "Exoplanets: Detection, Formation and Dynamics," in Exoplanets: Detection, formation and Dynamics, Y.-S. Sun, S. Fernaz-Mello, and J. –L. Zhou, Eds., *Proc. IAU Symposium* No. **IAU249.** pp. 17-24 (2008). [doi:10.1017/S174392130801630X]

28. Batalha, N. M, *et al.*, "Selection, prioritization, and characteristics of *Kepler* target stars," *ApJL* **713**, L109 (2010).

29. Brown, T. M., "Expected detection and false-alarm rates for transiting jovian planets," *ApJ* **593** L125-8 (2003).

30. Windemuth, D., Agol, E., Carter, J. et al., "An automated method to detect transiting circumbinary planets," *MNRS* **490**:1313-1324 (2019).

31. Gilliland, R. L., et al., "*Kepler* mission stellar and instrument noise properties," *Astrophys. J. Suppl.* **197**:6 (2011).

32. Basri, G., Walkowicz, L. M., Reiners, A., "Comparison of *Kepler* photometric variability with the Sun on different timescales" *APJ* **769**:37 (2013).

33. Catanzarite, J., and Shao, M., "The occurrence rate of Earth analog planets orbiting Sun-like stars," *ApJ* **738**:151 (2011).

34. Silbert, A., Gaidos, E., & Wu, Y., "The Occurrence of Potentially Habitable Planets Orbiting M Dwarfs Estimated from the Full *Kepler* Dataset and an Empirical Measurement of the Detection Sensitivity," *ApJ* **799**:180 (2015).

35. Petigura, E. A., Howard, A. W., & Marcy, G. W, "Prevalence of Earth-size planets orbiting Sun-like stars," *PNAS* 11019273P (2013).

36. Dressing, C. D., and Charbonneau, D., "The occurrence of potentially habitable planets orbiting M dwarfs estimated from the full *Kepler* dataset and an empirical measurement of the detection sensitivity," *ApJ* **807**:45 (2015).

37. Mulders, G. D., Pascucci, I., Apai, D, et al., "The exoplanet population simulator. I. The inner edges of planetary systems," *AJ* **156**:24 (2018).

38. Traub, W.A., "*Kepler* exoplanets: A new method of population analysis," arXiv:1605.02255 (2016).

**First Author** retired from NASA Ames in 2015, but continues as a NASA Ames Research Associate. He received BS and MS degrees in physics from the University of Wisconsin (Madison) in 1960 and 1962, respectively, and a MS in meteorology from California State University (San Jose, CA) in 1982. He is the author of more than 225 papers. During his 58 years at Ames, he contributed to: the design of the Apollo heat shield, modeling of the effects of nitrogen and chlorine pollutants on the Earth's ozone layer, lightning activity in planetary atmospheres, and modeling the electrical properties of the Titan atmosphere. He is a Co-Investigator on the Huygens Entry Probe and is the Principal Investigator for the *Kepler* Mission.

## <u>Caption List</u>
Fig. 1 Schematic diagram of point design of the *Kepler* instrument.
Fig. 2 Integrated spacecraft and photometer.
Fig. 3 Expected number of detections of planets in the HZ versus spectral type.
Fig. 4 Expected number of detections of planets not in the HZ versus spectral type.
Fig. 5 Mission capability of detecting planets of various size in the HZ versus the number of stars observed for a fixed FOV.

Figure 6. Model results for the values of the MF, number of small planets (Rp ≤ 1.3 R⊕) in the HZ, total number of planets (Rp ≤ 2 R⊕) in the HZ, and the number of non-HZ planets versus the number of monitored stars.

Figure 7. Expected number of detections based on the instrument capability, characterization of the stars in the *Kepler* FOV, and the assumptions of one terrestrial planet in each HZ and a requirement of detecting a minimum of 4 transits. Results in the left-hand panel are for the selection of 170,000 brightest stars (Case#1). Right-hand panel shows results for 170,000 "cherry-picked" stars (Case#2).

Figure 8. Expected number of detections for non-HZ planets distributed over a range of semi-major axes for a mission duration of 4 years.

Figure 9. Mission capability to detect HZ-planets when the mission duration is reduced from 4 years to 1 year for cherry-picked stars (Case#2). (Left to right and then down; 4, 3, 2, & 1 year mission durations.)

Figure 10. Fraction of the population of small planets (Rp ≤1.3 R⊕) in the HZ that can be detected in the FOV. Black curves represent Case#1 while red curves represent Case#2.

Figure 11. Predicted number of detectable small planets (Rp ≤ 1.3 R⊕) versus the value of the recognition threshold. Note that a value of 7 was chosen for Mission point design and the calculations in this paper.

Table 1 MF Value, Number of Detectable Planets in the HZ, and Number of Detectable Planets not in the HZ.

**Appendix 1. Outline of Merit Function Requirements and Design**

Requirements
1) Provide a numerical value proportional to the science value of the expected results.
2) Normalize the score such that the total value will be 100 points when all the requirements stated in the CSR are met.
3) Evaluate planets in the Habitable Zone (HZ-planets), and planets not in the HZ ("non-HZ planets") separately. Assign more weight to the first group and less to the second in alignment with their importance to the goals of the mission. For the *Kepler* Mission, the score for HZ-planets received a weight of 0.65 while the non-HZ planets received a weight of 0.35.
4) Assign values for subsets of the goals and weight their importance. For the *Kepler* Mission, the weights for the capability to detect a small planets in the HZ were set much higher than for larger planets because that choice ensured that all larger planets could also be detected. The score was set higher for planets with large signal-to-noise ratios (SNR) transit patterns because they produce a higher recognition probability and a lower false-positive probability. For planets not in the HZ, the score for planets orbiting close to their star is less than those at larger semi-major axis to ensure a more uniform determination of the occurrence frequency vs semi-major axis.
5) The inputs should include instrument and mission parameters that affect the model results. For example: a) photon flux (which is affected by the aperture, transmission and bandwidth of optics and filters, and the efficiency of the detectors.); b) mission duration; c) and size and sky location of the FOV.
6) The score for expected discoveries that do not make a substantial contribution to the accomplishment of mission goals or which are accomplished irrespective of mission choices should be set to zero to force the mission emphasis on the stated goals.

Outline of Merit Function Design
Set Constants and Evaluate Input Tables
1) Mass, size, and occurrence frequency of each star type
2) Visual magnitudes to be considered: 9-16.5
3) Frequency/amplitude of stellar/solar variability
4) Photometer, read, and jitter noise values from CSR Table F5

Two components are considered;
1) "Planets in HZ": Range of planet sizes in habitable zone (HZ) of each star type
2) "Non-HZ" planets: all planets are $1R_\oplus$ in size and distributed with logarithm of semi-major axis with values near 1 AU regardless of star type.

COMPONENT #1. Planets in HZ
Input:
- Planet sizes (relative to Earth). Mars, Earth, 1.3 Earth, 2 Earth
- Score values for MF
    40 pts for each recognized planet in the HZ for planet radius $Rp = 0.53R_\oplus$
    20 pts for each recognized planet in the HZ for $Rp= 1\ R_\oplus$

       5 pts for each additional planet with Rp = 1.3 R$_\oplus$
       1 pts for each additional planet with Rp = 2 R$_\oplus$

Calculations:
- Calculate the number of transits/star for 4 & 6 year mission durations as a function of stellar type and semi-major axis
- Calculate the SNR for 4yr and 6 year mission durations for all magnitudes, spectral types, & planet sizes
- Calculate the probability of a transit for each star type in the HZ
- Count the number of stars that have the correct orbital alignment vs spectral type and apparent magnitude and count the number of stars that have 3 or more transits and a detection threshold value of 7$\sigma$ or greater. Include a correction for the recognition rate as function of threshold value.
- Compute the score for the chosen parameters

COMPONENT #2. Non-HZ planets
Procedure is similar to Component #1 with the following exceptions:
- The size of all planets = 1 R$_\oplus$
- There exists one planet at each of nine semi-major axe, but do not count planets that have a value of the semi-major axis that duplicates one in the HZ of that star type.
- Multiply the results by the expected number of non-HZ planets divided by the 8 possible positions.

***Disclosures***
No conflicts of interest